\documentclass{article}
\usepackage{ijcai25}
\usepackage{times}
\usepackage{soul}
\usepackage{tikz}
\usetikzlibrary{shapes.geometric, arrows, positioning,fit,calc}

\usepackage[edges]{forest}
\definecolor{hidden-draw}{RGB}{0,0,0}
\definecolor{hidden-pink}{rgb}{0.98, 0.94, 0.75}
\definecolor{level3}{RGB}{216,226,234}
\definecolor{level1}{rgb}{0.882, 0.835, 0.906}
\definecolor{level2}{rgb}{0.8, 0.8, 1.0}
\definecolor{level0}{rgb}{1.0, 0.71, 0.76}
\definecolor{level4}{rgb}{0.98, 0.92, 0.84}

\definecolor{myred}{RGB}{247,226,231}
\definecolor{myblue}{RGB}{216,226,234}
\definecolor{myyellow}{RGB}{252,238,221}
\definecolor{mypurple}{RGB}{233,229,241}

\usepackage{url}
\usepackage[hidelinks]{hyperref}
\usepackage[utf8]{inputenc}
\usepackage[small]{caption}
\usepackage{graphicx}
\usepackage{amsthm}
\usepackage{amsmath,amssymb,amsfonts}
\usepackage{booktabs}
\usepackage{subcaption}
\usepackage{algorithm}
\usepackage{algorithmic}
\usepackage{multirow}
\usepackage{tabularx}
\usepackage{enumitem}
\usepackage{bbm}
\usepackage{balance}
\usepackage[flushleft]{threeparttable}

\usepackage{color, colortbl}
\definecolor{Gray}{gray}{0.9}
\usepackage{setspace}
\usepackage{forest}
\usepackage{titlesec}
\titlespacing*{\paragraph}{0pt}{0.7ex}{0.8ex} 
\title{A Survey on Video Analytics in Cloud-Edge-Terminal Collaborative Systems}
\author{
Linxiao Gong$^{1,3}$\and
Hao Yang$^1$\and
Gaoyun Fang$^4$\and
Bobo Ju$^1$\and
Juncen Guo$^1$\and\\
Xiaoguang Zhu$^5$\and
Xiping Hu$^6$\and
Yan Wang$^1$\and
Peng Sun$^2$\and
Azzedine Boukerche$^7$\\
\affiliations
$^1$Fudan University,\quad
$^2$Duke Kunshan University,\quad
$^3$University of Toronto,\quad\\
$^4$Imperial College London,\quad
$^5$University of California, Davis,\quad
$^6$SMBU,\quad
$^7$University of Ottawa\quad
\emails 
\{lxgong21, haoyang24, bbju21, guojc23, yanwang19\}@fudan.edu.cn, p.fang23@imperial.ac.uk, xgzhu@ucdavis.edu, huxp@smbu.edu.cn, peng.sun568@duke.edu, aboukerc@uottawa.ca
}
\begin{document}
\maketitle
\begin{abstract}
The explosive growth of video data has driven the development of distributed video analytics in cloud-edge-terminal collaborative (CETC) systems, enabling efficient video processing, real-time inference, and privacy-preserving analysis. Among multiple advantages, CETC systems can distribute video processing tasks and enable adaptive analytics across cloud, edge, and terminal devices, leading to breakthroughs in video surveillance, autonomous driving, and smart cities. In this survey, we first analyze fundamental architectural components, including hierarchical, distributed, and hybrid frameworks, alongside edge computing platforms and resource management mechanisms. Building upon these foundations, edge-centric approaches emphasize on-device processing, edge-assisted offloading, and edge intelligence, while cloud-centric methods leverage powerful computational capabilities for complex video understanding and model training. Our investigation also covers hybrid video analytics incorporating adaptive task offloading and resource-aware scheduling techniques that optimize performance across the entire system. Beyond conventional approaches, recent advances in large language models and multimodal integration reveal both opportunities and challenges in platform scalability, data protection, and system reliability. Future directions also encompass explainable systems, efficient processing mechanisms, and advanced video analytics, offering valuable insights for researchers and practitioners in this dynamic field.
\end{abstract}
\section{Introduction}\label{sec1}
\begin{figure}[t]
    \centering
    \includegraphics[width=0.48\textwidth]{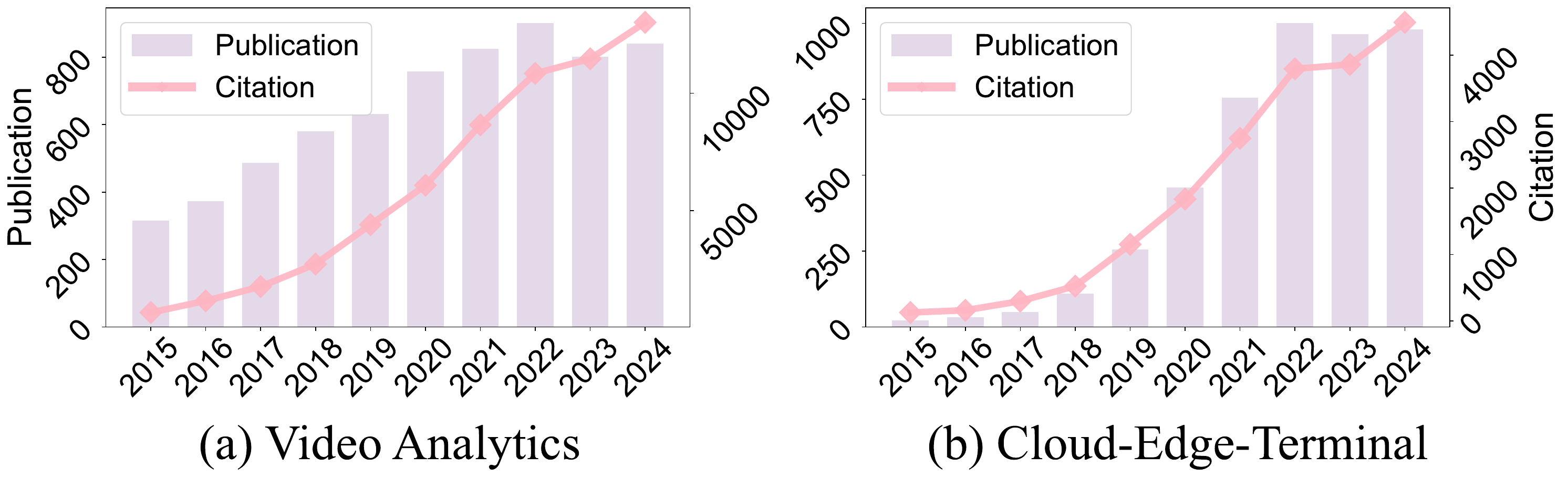}
    \vspace{-20px}
    \caption{Publication and citation statistics over the last decade.}
    \label{fig:pub_stats}
    \vspace{-15px}
\end{figure}

Widespread deployment of surveillance cameras, smartphones, and IoT devices has generated unprecedented volumes of video data, driving innovations across traffic monitoring, smart cities, and industrial automation~\cite{zhang2021serverless}. Video analytics has become increasingly critical, driven by the growing need for real-time insights across diverse scenarios~\cite{lu2024samedge}, from predictive maintenance systems~\cite{ji2024task} and autonomous vehicles demanding instantaneous decisions~\cite{zhu2024edge}, to advanced interactive applications like augmented and virtual reality platforms requiring responsive video analysis for compelling user experiences~\cite{zhou2024survey}. Consequently, cloud-edge-terminal collaborative (CETC) systems have emerged as a promising solution to manage this computational landscape by combining cloud servers with edge and terminal devices~\cite{gao2024edgevision}, while resource management strategies adapt to varying content complexity and processing demands~\cite{lu2023turbo}, as evidenced by the growing publication and citation trends, shown in Fig.~\ref{fig:pub_stats}.

Modern cloud-based video analytics, while leveraging substantial computational resources, encounters fundamental operational challenges in bandwidth utilization, response latency, and network reliability~\cite{gu2024aienhanced}. Specifically, large-scale video data transmission to centralized servers creates significant network bottlenecks, especially evident in deployments with multiple high-resolution camera streams. Such network congestion directly impacts real-time applications like traffic monitoring and autonomous driving systems, where processing delays can critically affect decision-making capabilities~\cite{ji2024task}. Moreover, the dependence on consistent cloud connectivity poses significant challenges for system deployment in areas with unstable or constrained network infrastructure~\cite{vo2022edge}, thereby affecting the overall system reliability and effectiveness.

Although existing surveys have studied video analytics and edge-cloud computing, there is still a lack of systematic analysis of video analytics in CETC systems. Previous surveys either focus on specific aspects like edge computing \cite{zhu2024edge}, video streaming \cite{zhou2024survey}, or generative AI \cite{xu2024unleashing}, without systematically examining the integration of video analytics across cloud, edge, and terminal tiers. While \cite{gu2024aienhanced} provides insights into AI-enhanced cloud-edge-terminal networks, it does not specifically address the unique challenges and opportunities in video analytics applications. We aim to bridge this gap by providing a systematic review of video analytics in CETC systems, with particular attention to the interplay between different tiers and their roles in efficient video processing and analysis. The taxonomy of this survey is illustrated in Fig.~\ref{fig:taxonomy}, which categorizes existing works into edge-centric, cloud-centric, and hybrid video analytics. The main contributions of this survey are fourfold: \textbf{1)} \textit{A systematic taxonomy}. To the best of our knowledge, we are the first to present a structured taxonomy that categorizes existing works in video analytics across cloud, edge, and terminal tiers, covering architectural paradigms, processing strategies, and optimization techniques. \textbf{2)} \textit{A comprehensive review}. We provide an analysis of current research progress in edge-centric, cloud-centric, and hybrid video analytics, examining their advantages, limitations, and practical implementations. \textbf{3)} \textit{Novel insights}. We identify emerging trends and challenges in CETC video analytics, including platform integration, system scalability, data protection, and the integration of advanced technologies like LLMs and multimodal learning. \textbf{4)} \textit{Future directions}. We discuss potential research opportunities and outline promising directions for advancing video analytics in CETC systems.
\tikzstyle{my-box}=[
    rectangle,
    draw=hidden-draw,
    rounded corners,
    text opacity=1,
    minimum height=1.5em,
    minimum width=5em,
    align=center,
    fill opacity=.5,
    line width=0.8pt,
    font=\fontsize{11}{15}\selectfont,
    text=black,
]

\tikzstyle{tklevel0}=[my-box,
    fill=level0!50,
    font=\LARGE,
]

\tikzstyle{tklevel1}=[my-box,
    fill=level1,
    font=\LARGE,
]

\tikzstyle{tklevel2}=[my-box,
    fill=level2!60,
]
\tikzstyle{tklevel3}=[my-box,
    fill=level3!60,
]

\tikzstyle{leaf}=[my-box,
    fill=level4!50,
]

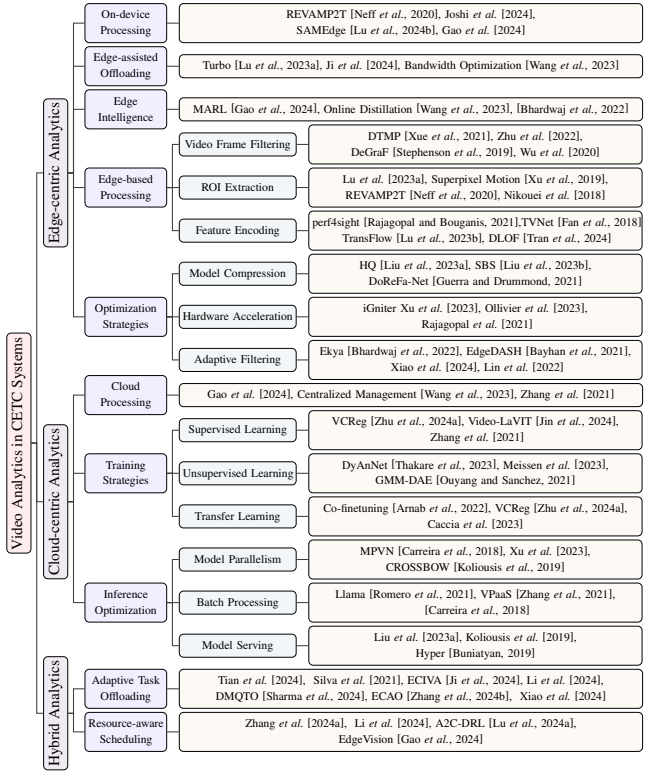
\begin{figure}[t]
    \centering
    \resizebox{0.48\textwidth}{!}
    {
        \begin{forest}
            forked edges,
            for tree={
                fill=level0!50,
                grow=east,
                reversed=true,
                anchor=base west,
                parent anchor=east,
                child anchor=west,
                base=center,
                font=\large,
                rectangle,
                draw=hidden-draw,
                rounded corners,
                align=center,
                minimum width=4em,
                edge+={darkgray, line width=1pt},
                s sep=3pt,
                line width=0.8pt,
                ver/.style={rotate=90, child anchor=north, parent anchor=south, anchor=center},
            },
            where level=0{ver,tklevel0}{},
            where level=1{ver,tklevel1}{},
            where level=2{text width=7em,tklevel2,}{},
            where level=3{text width=10.3em,tklevel3,}{},
            [Video Analytics in CETC Systems
                [Edge-centric Analytics
                    [On-device\\Processing,align=center
                        [REVAMP2T \cite{neff2020revamp2t}{,}~Joshi \textit{et al.}~\shortcite{joshi2024integration}{,}\\SAMEdge \cite{lu2024samedge}{,}~Gao \textit{et al.}~\shortcite{gao2024edgevision}, leaf,text width=42.8em,]
                    ]
                    [Edge-assisted\\Offloading,align=center
                        [Turbo \cite{lu2023turbo}{,}~Ji \textit{et al.}~\shortcite{ji2024task}{,} Bandwidth Optimization \cite{wang2023shoggoth}, leaf,text width=42.8em,]
                    ]
                    [Edge\\Intelligence
                        [MARL \cite{gao2024edgevision}{,} Online Distillation \cite{wang2023shoggoth}{,} \cite{bhardwaj2022ekya}, leaf,text width=42.8em,]
                    ]
                    [Edge-based\\Processing
                        [Video Frame Filtering
                            [DTMP \cite{xue2021denoisingbased}{,} Zhu \textit{et al.}~\shortcite{zhu2022open}{,}\\ DeGraF \cite{stephenson2019degrafflow}{,}~Wu \textit{et al.}~\shortcite{wu2020descriptive}, leaf,text width=30.6em,]
                        ]
                        [ROI Extraction
                            [Lu \textit{et al.}~\shortcite{lu2023turbo}{,} Superpixel Motion \cite{xu2019robust}{,}\\
                            REVAMP2T \cite{neff2020revamp2t}{,}~Nikouei \textit{et al.}~\shortcite{nikouei2018realtime}, leaf,text width=30.6em,]
                        ]
                        [Feature Encoding
                            [perf4sight~\cite{rajagopal2021perf4sight}{,}TVNet~\cite{fan2018endtoend}
                            \\TransFlow \cite{lu2023transflow}{,} DLOF \cite{tran2024deeplearning}, leaf,text width=30.6em,]
                        ]
                    ]
                    [Optimization\\Strategies
                        [Model Compression
                            [HQ \cite{liu2023hyperspherical}{,} SBS \cite{liu2023singlepath}{,}\\DoReFa-Net \cite{guerra2021automatic}, leaf,text width=30.6em,]
                        ]
                        [Hardware Acceleration
                            [iGniter Xu \textit{et al.}~\shortcite{xu2023igniter}{,} Ollivier \textit{et al.}~\shortcite{ollivier2023sustainable}{,}\\~Rajagopal \textit{et al.}~\shortcite{rajagopal2021perf4sight}, leaf,text width=30.6em,]
                        ]
                        [Adaptive Filtering 
                            [Ekya \cite{bhardwaj2022ekya}{,} EdgeDASH \cite{bayhan2021edgedash}{,}\\~Xiao \textit{et al.}~\shortcite{xiao2024transcodingenabled}{,}~Lin \textit{et al.}~\shortcite{lin2022efficient}, leaf,text width=30.6em,]
                        ]
                    ]
                ]
                [Cloud-centric Analytics
                    [Cloud\\Processing
                        [Gao \textit{et al.}~\shortcite{gao2024edgevision}{,} Centralized Management \cite{wang2023shoggoth}{,}~Zhang \textit{et al.}~\shortcite{zhang2021serverless}, leaf,text width=42.8em,]
                    ]
                    [Training\\Strategies
                        [Supervised Learning
                            [VCReg \cite{zhu2024variancecovariance}{,} Video-LaVIT \cite{jin2024videolavit}{,}\\~Zhang \textit{et al.}~\shortcite{zhang2021serverless}, leaf,text width=30.6em,]
                        ]
                        [Unsupervised Learning
                            [DyAnNet \cite{thakare2023dyannet}{,}  Meissen \textit{et al.}~\shortcite{meissen2023unsupervised}{,}\\GMM-DAE~\cite{ouyang2021video}, leaf,text width=30.6em,]
                        ]
                        [Transfer Learning
                            [Co-finetuning \cite{arnab2022transfer}{,} VCReg \cite{zhu2024variancecovariance}{,}\\ Caccia \textit{et al.}~\shortcite{caccia2023computeoptimal}, leaf,text width=30.6em,]
                        ]
                    ]
                    [Inference\\Optimization
                        [Model Parallelism
                            [MPVN \cite{carreira2018massively}{,}~Xu \textit{et al.}~\shortcite{xu2023igniter}{,}\\CROSSBOW \cite{koliousis2019crossbow}, leaf,text width=30.6em,]
                        ]
                        [Batch Processing
                            [Llama \cite{romero2021llama}{,} VPaaS~\cite{zhang2021serverless}{,}\\~\cite{carreira2018massively}, leaf,text width=30.6em,]
                        ]
                        [Model Serving
                            [Liu \textit{et al.}~\shortcite{liu2023hyperspherical}{,} Koliousis \textit{et al.}~\shortcite{koliousis2019crossbow}{,}\\ Hyper~\cite{buniatyan2019hyper}, leaf,text width=30.6em,]
                        ]
                    ]
                ]
                [Hybrid Analytics,align=center,
                    [Adaptive Task\\Offloading
                        [Tian \textit{et al.}~\shortcite{tian2024dynamic}{,} ~Silva \textit{et al.}~\shortcite{silva2021energyaware}{,} ECIVA~\cite{ji2024task}{,}~Li \textit{et al.}~\shortcite{li2024twotimescale}{,}\\ DMQTO~\cite{sharma2024deep}{,} ECAO~\cite{zhang2024energy}{,} ~Xiao \textit{et al.}~\shortcite{xiao2024transcodingenabled},leaf,text width=42.8em,]
                    ]
                    [Resource-aware\\Scheduling
                        [Zhang \textit{et al.}~\shortcite{zhang2024hybrid}{,} ~Li \textit{et al.}~\shortcite{li2024twotimescale}{,} A2C-DRL~\cite{lu2024a2cdrl}{,}\\EdgeVision~\cite{gao2024edgevision},leaf,text width=42.8em,]
                    ]
                ]
            ]
        \end{forest}
    }    
\vspace{-16px}
\caption{Taxonomy of video analytics in CETC systems.}
\label{fig:taxonomy}
\vspace{-15pt}
\end{figure}

\section{Foundation of CETC}\label{sec2}
\subsection{Architectural Paradigms}\label{sec2.1}
Architectures in CETC systems adopt three core paradigms differentiated by resource coordination mechanisms.
\paragraph{Hierarchical Architectures.}\label{sec2.1.1}
Hierarchical architectures form the backbone of contemporary CETC video analytics systems, enabling distributed processing across multiple computational tiers~\cite{joshi2024integration}. Within this framework, the cloud serves as the central orchestrator, managing complex computations while edge devices positioned near data sources handle intermediate computations and feature extraction. Specifically, given raw video data $V_t$ originates from terminal devices, undergoing initial processing $f_t$ locally before edge-level transformations $f_e$ generate processed data $D_e$ for cloud analysis $f_c$ and subsequent control signal $C$ generation. Resource optimization occurs through strategic task allocation $t_i$, which minimizes latency $L_j$ based on available computational resources $R_j$ across deployment tiers~\cite{tian2024dynamic}. Edge-level filtering and aggregation mechanisms effectively reduce data volumes, while centralized management streamlines system control and enhances security through restricted raw data access~\cite{vo2022edge}. Notable challenges include potential vulnerability to cloud-level failures and network disruptions~\cite{sedlak2024equilibrium}.
\paragraph{Distributed Architectures.}\label{sec2.1.2}
Collaborative processing paradigms form the foundation of modern distributed video analytics, enabling autonomous edge and terminal device cooperation while reducing cloud dependencies. Fundamental to this approach is the implementation of federated learning~\cite{guo2023fedfsnet}, which aggregates local model updates $w_i^t$ from $N$ participating devices into a global model $W_t = \frac{1}{N} \sum_{i=1}^{N} w_i^t$ without raw data exchange. Resource utilization optimization involves careful consideration of task offloading decisions through utility functions $\min \sum_{i=1}^{N} \sum_{j=1}^{M} x_{ij} (c_{ij} + l_{ij})$ with task-to-tier allocation variable $x_{ij}$, balancing communication costs $c_{ij}$ and execution costs $l_{ij}$ under system constraints~\cite{ji2024task}. Advanced scheduling mechanisms~\cite{kuswiradyo2024optimizing} complement these efforts through dynamic resource allocation. Additionally, the computing continuum paradigm further bridges edge-cloud tiers for adaptive resource sharing and workload distribution, while WebAssembly~\cite{sedlak2024equilibrium} enables cross-platform compatibility through unified runtime environments.

\paragraph{Hybrid Architectures.}\label{sec2.1.3}
Adaptive resource allocation mechanisms characterize hybrid architectures, enabling dynamic workload distribution across cloud, edge, and terminal tiers based on operational parameters~\cite{bayhan2021edgedash}. To optimize this distribution, a cost-minimization framework guides task execution decisions, where $C_i(R_j)$ represents the resource-specific cost for task $T_i$ on computing unit $R_j$. Within this framework, communication efficiency optimization focuses on minimizing latency $L_i$, expressed as $D_i/B_j$ for data volume $D_i$ and available bandwidth $B_j$~\cite{silva2021energyaware}. System performance benefits from content-aware video encoding strategies that adjust rates $E_i$ based on bandwidth availability and content complexity~\cite{zhang2021serverless}. Additionally, hierarchical learning, with initial training in the cloud and subsequent fine-tuning at the edge, enables efficient model updates \cite{xu2024unleashing}. To enhance computational efficiency, transfer learning approaches facilitate model adaptation through cloud-initialized training refined at edge nodes~\cite{gao2024edgevision}. Finally, performance optimization leverages resource-aware scheduling algorithms that consider device-specific capabilities.

\subsection{Edge Computing Platforms}\label{sec2.2.2}
Modern platforms including NVIDIA Jetson and AWS Wavelength leverage containerization technologies (e.g., Docker and Kubernetes) for streamlined deployment~\cite{lu2024samedge}, with selection criteria driven by computational and power requirements~\cite{bhardwaj2022ekya}. Resource-optimized computing platforms enable efficient deployment of video analytics models on constrained devices like single-board computers and embedded systems.  By applying queueing theory principles, system architects model resource allocation across $N_s$ identical servers with arrival rate $\lambda$ and service rate $\mu$, maintaining system utilization $\rho = \frac{\lambda}{N_s\mu}$ below unity to ensure processing stability~\cite{ji2024task}. Real-time performance metrics employ Little's Law, where average waiting time $W = \frac{q_l}{\lambda}$ reflects queue length $q_l$~\cite{gao2024edgevision}. Likewise, Hardware acceleration with graphics processing units (GPUs) and field-programmable gate arrays (FPGAs) offers significant performance improvements, often quantified by the speedup ratio $S = \frac{T_{\text{CPU}}}{T_{\text{accelerator}}}$, while power optimization necessitates a careful balance between static and dynamic power consumption $P = P_{\text{static}} + P_{\text{dynamic}}$.

\subsection{Resource Management}\label{sec2.2.4}
Resource management in CETC systems presents multifaceted challenges of optimizing computational efficiency across distributed infrastructure tiers.
\paragraph{Latency Minimization.} Resource allocation in CETC video analytics systems efficiently distributes computing, storage, and network resources, aiming to minimize system latency $L_{sys}$, which encompasses processing $L_{proc}$, transmission $L_{trans}$, and queuing $L_{queue}$ latencies, i.e., $L_{sys} = L_{proc} + L_{trans} + L_{queue}$. Recent works demonstrate innovative approaches to optimize each component: 1) For $L_{trans}$, \cite{zhang2021serverless} proposes a cloud-fog video streaming protocol that reduces bandwidth usage by 21\% through adaptive quality selection and fog-assisted correction; 2) $L_{queue}$ is addressed by A2C-DRL \cite{lu2024a2cdrl} that dynamically schedules tasks across edge-cloud nodes, achieving RTT reduction through decentralized reinforcement learning.

\paragraph{Energy Efficiency and Throughput Maximization.} Minimizing energy consumption $E_{total}$ is crucial, particularly at the edge and terminal devices. The total energy consumption across $N_d$ devices is calculated by summing individual device energy $E_i$ based on allocated computing resources $r_i$ and processing time $t_i$: $E_{total} = \sum_{i=1}^{N_d} E_i(r_i, t_i)$. \cite{hua2024energyefficient} explores energy-efficient allocation in heterogeneous edge-cloud computing, considering user mobility. In parallel, network throughput $T$ is maximized through efficient bandwidth allocation, computed as the sum product of allocated bandwidth $B_{jk}$ and achieved data rate $R_{jk}$ across $M$ links and $K$ users: $T = \sum_{j=1}^{M} \sum_{k=1}^{K} B_{jk} \cdot R_{jk}$. Furthermore, continuous learning of video analytics models on edge servers can optimize resource utilization and improve overall network performance \cite{bhardwaj2022ekya}.

\paragraph{Quality of Experience and Dynamic Adaptation.}  Quality of experience (QoE), often represented by a utility function $U(q)$, is a critical factor in video analytics. For example, EdgeDASH \cite{bayhan2021edgedash} framework explores network-assisted adaptive video streaming for edge caching, optimizing QoE by facilitating the use of cached video content and improving cache hit rates and video quality. Reinforcement learning (RL) offers a promising approach, with \cite{ji2024task} introducing a GRU-enhanced deep Q-Network (GEDQN) for task offloading. Building upon this foundation, hybrid approaches combining deep reinforcement learning with multi-objective optimization demonstrate superior performance in balancing task completion time, cost, energy consumption, and system reliability \cite{zhang2024hybrid}. Furthermore, opportunistic data enhancement using idle GPU resources \cite{lu2023turbo}, showcases innovative strategies for resource utilization optimization.
\section{Edge-centric Video Analytics}\label{sec3}
\subsection{On-device Processing}\label{sec3.2}
On-device video analytics in CETC systems offers advantages such as reduced latency for real-time applications like autonomous driving and security surveillance \cite{neff2020revamp2t}, and minimized bandwidth consumption, particularly crucial in bandwidth-constrained environments \cite{joshi2024integration}.  Additionally, on-device processing enhances privacy by localizing sensitive video data and improves system scalability and resilience through distributed workload processing \cite{gao2024edgevision}. However, challenges include limited computational resources and power budgets on edge devices, restricting model complexity and performance \cite{sedlak2024equilibrium}.  The heterogeneity of edge devices also introduces integration and management complexities, while ensuring reliability and robustness against device failures or network disruptions remains a concern \cite{vo2022edge}.  Optimizing model size and computational efficiency for resource-constrained devices while maintaining accuracy is crucial, necessitating techniques like model compression, hardware acceleration, and efficient task offloading strategies \cite{ji2024task}.  Some works like REVAMP$^2$T \cite{neff2020revamp2t} and SAMEdge \cite{lu2024samedge} demonstrate the feasibility of on-device processing for real-time, privacy-aware pedestrian tracking and supporting complex models on edge devices.
\subsection{Edge-assisted Offloading}\label{sec3.3}
Edge servers are crucial for pre-processing and filtering video data in CETC systems \cite{kuswiradyo2024optimizing}, reducing the volume of data transmitted to the cloud and thus minimizing bandwidth consumption and enhancing system efficiency.  Specifically, they perform initial processing, such as Turbo \cite{lu2023turbo} and ROI extraction \cite{ji2024task}, to reduce redundancy and extract key information before cloud transmission.  Such edge pre-processing contributes to lower latency by performing computations closer to the data source \cite{buniatyan2019hyper}, enabling near real-time analytics even with limited bandwidth \cite{wang2023shoggoth}.  Additionally, edge-assisted offloading improves resource utilization by distributing the computational workload, optimizing the use of both edge and cloud resources for more efficient and scalable video analytics systems \cite{hua2024energyefficient}.
\subsection{Edge Intelligence}\label{sec3.4}
Advanced computational methodologies in CETC systems leverage intelligent decision-making capabilities at network edges, enabling real-time insights with minimal cloud dependencies~\cite{sharma2024deep}. Specifically, online knowledge distillation techniques effectively adapt to dynamic scenes while mitigating data drift, complemented by adaptive training approaches utilizing small batches and selective frame sampling~\cite{wang2023shoggoth}. Integrating AI at the network edge successfully addresses stability concerns and device overload while optimizing energy-latency trade-offs, enhanced by incremental learning with real-time data streams~\cite{joshi2024integration}. In addition, multiagent reinforcement learning (MARL) like EdgeVision~\cite{gao2024edgevision} also facilitate autonomous learning across edge nodes, while direct deployment allows for continuous learning and adaptation of video analytics models on edge servers, addressing the challenges of data drift and ensuring long-term model accuracy \cite{bhardwaj2022ekya}.

\subsection{Edge-based Video Processing}\label{sec3.5}
Edge computing plays a crucial role in CETC systems by enabling video processing near the data source. This section explores key techniques for edge-based video processing.
\paragraph{Video Frame Filtering.}\label{sec3.5.1}

Edge video frame filtering (VFF) optimizes computational efficiency by discarding redundant frames and extracting key information through intelligent frame selection and background modeling. Keyframe extraction retains informative frames while background subtraction isolates moving objects for downstream analysis \cite{wu2020descriptive}. Advanced background modeling techniques, from superpixel motion detection \cite{xu2019robust} to sparse representation with outlier removal \cite{tran2024deeplearning}, enable robust foreground segmentation under dynamic conditions. Recent innovations like weakly supervised OpenVAD which integrates evidential deep learning and normalizing flows \cite{zhu2022open}, enhance robustness in challenging scenarios by effectively identifying both known and novel anomalies.  Additionally, denoising-based turbo message passing (DTMP) \cite{xue2021denoisingbased} optimizes background subtraction in compressed video streams, while CNNs provide efficient filtering for depth videos, particularly in complex lighting conditions. DeGraF \cite{stephenson2019degrafflow} further enhances optical flow estimation by offering uniformly distributed and tunable feature points. Consequently, by leveraging lightweight models and employing model selection strategies at the edge, real-time video analytics can be achieved even with limited resources and bandwidth \cite{gao2024edgevision}.

\paragraph{Region of Interest Extraction.}\label{sec3.5.2}
ROIE is crucial for edge-centric video analytics, identifying relevant areas within video frames at the edge to reduce computational burden and bandwidth requirements for downstream processing.  Efficient ROIE is paramount in edge-centric video analytics, employing diverse techniques like object detection~\cite{lu2023turbo} can identify informative regions for subsequent processing.  Additionally, motion detection algorithms, such as those employed in the blockchain-enabled surveillance video indexing scheme \cite{nikouei2018realtime}, effectively identify regions exhibiting significant motion, enabling efficient extraction of motion information and object segmentation. Superpixel-based motion detection \cite{xu2019robust} combines spatial and temporal information for robust extraction under challenging conditions. Interactive segmentation techniques, such as REVAMP$^2$T \cite{neff2020revamp2t}, enable users to refine ROI selection through clicks, thereby enhancing the accuracy and efficiency of video analysis tasks.

\paragraph{Feature Encoding.}\label{sec3.5.3}
Edge computing layers perform critical visual abstraction through compact feature extraction, where convolutional networks like perf4sight~\cite{rajagopal2021perf4sight} establish hierarchical spatial representations. Temporal dynamics are captured via optical flow estimation techniques, with DeGraF-Flow~\cite{stephenson2019degrafflow} leveraging dense gradient fields and TransFlow~\cite{lu2023transflow} introducing transformer-based global dependency modeling. Complementing these approaches, TVNet~\cite{fan2018endtoend} and DLOF~\cite{tran2024deeplearning} employ deep architectures to learn flow-like motion patterns for enhanced video understanding. Additionally, feature encoding exploits temporal context through two-stream architectures \cite{ji2024task} and self-supervised contrastive learning \cite{noguchi2023egovehicle} to enhance video analytics performance.

\subsection{Optimization Strategies}\label{sec3.6}
Edge video analytics optimization strategies encompass model compression, hardware acceleration, and adaptive filtering for computational efficiency.
\paragraph{Model Compression.}\label{sec3.6.1}
Optimizing deep neural networks for edge deployment necessitates innovative approaches to reduce computational overhead while maintaining analytical accuracy through pruning and quantization \cite{joshi2024integration}. Structured pruning eliminates redundant filters or channels to enhance inference speed on generic hardware, whereas unstructured pruning targets sparse weight removal for potential accuracy preservation at the cost of specialized accelerator requirements. Quantization reduces the precision of model parameters and activations, representing them with fewer bits, as exemplified by binarized neural networks and DoReFa-Net \cite{guerra2021automatic}, drastically reducing model size and computational cost. Hyperspherical quantization methods like hyperspherical quantization (HQ) \cite{liu2023hyperspherical} improve the accuracy of quantized models by minimizing the distance between full-precision and quantized weights. Joint pruning and quantization techniques like single-path bit sharing (SBS) \cite{liu2023singlepath}, offer a unified approach for higher compression ratios. To optimize memory usage during training, techniques like entropy-based pruning can be employed to minimize the size of intermediate activations, making it possible to train larger models or utilize devices with limited memory capacity \cite{caccia2023computeoptimal}.

\paragraph{Hardware Acceleration.}\label{sec3.6.2} 
Edge hardware acceleration utilizes specialized hardware such as GPUs \cite{xu2023igniter} and FPGAs \cite{ollivier2023sustainable} to improve the efficiency of video processing at the edge. The utilization of GPUs in edge computing environments, as demonstrated by the perf4sight \cite{rajagopal2021perf4sight}, highlights their effectiveness in accelerating computationally intensive tasks such as CNN-based object detection and video analysis.  In contrast, FPGAs provide flexibility and customizability, enabling the implementation of optimized hardware architectures for specific video processing tasks and achieving high performance and energy efficiency.  Their reconfigurability also allows adaptation to evolving video analytics algorithms and application requirements.  In heterogeneous embedded systems, the use of such specialized hardware further enhances performance and energy efficiency \cite{lin2022efficient}.  Considering sustainable AI, evaluating hardware acceleration options based on factors like embodied energy and carbon footprint becomes crucial, with technologies like Racetrack memory PIM offering potential advantages \cite{ollivier2023sustainable}.

\paragraph{Adaptive Filtering.}\label{sec3.6.3}
Adaptive filtering in edge-centric video analytics dynamically adjusts filtering parameters based on both video content characteristics and real-time edge resource availability to optimize performance in resource-constrained environments.  For example, in complex scenes or when motion is high, more intensive filtering may be necessary for effective feature extraction, while under simpler content or low resource availability, less intensive filtering can conserve power and minimize latency. Such dynamic adaptation aligns with opportunistic resource utilization \cite{lu2023turbo}, and resonates with quality adaptation mechanisms in adaptive video streaming (e.g., EdgeDASH \cite{bayhan2021edgedash}).  Adaptive filtering can also optimize VFF and ROIE by dynamically selecting keyframes, adjusting background subtraction parameters, or refining object/motion detection parameters based on video content dynamics.  Systems like Ekya \cite{bhardwaj2022ekya} leverage this approach to balance retraining and inference accuracy on edge servers, complementing resource-aware scheduling \cite{xiao2024transcodingenabled}.  In addition, parameter selection can be informed by video content variability for consistent quality, and the system in \cite{lin2022efficient} demonstrates the importance of optimizing data flows across heterogeneous compute units, which can further enhance adaptive filtering's effectiveness.

\section{Cloud-centric Video Analytics}\label{sec4}
\subsection{Cloud-based Processing}\label{sec4.2}
Powerful computational capabilities transform complex video analytics through cloud-based infrastructures, leveraging vast computational resources and storage capacity \cite{gao2024edgevision}.  The abundant computing power enables execution of computationally intensive deep learning models for tasks like object detection and action recognition, crucial for handling high-resolution video and complex analytical pipelines.  Correspondingly, the virtually unlimited storage accommodates large datasets generated by video analytics, facilitating long-term trend analysis and pattern recognition. Additionally, centralized cloud resources facilitate training sophisticated video analytics models using large datasets, leading to improved accuracy and generalization through training paradigms including supervised, unsupervised, and transfer learning \cite{zhang2021serverless}. Cloud platforms also offer scalability and flexibility, adapting to fluctuating workloads and varying processing demands, essential for supporting real-time video processing from multiple sources \cite{lu2023turbo}.  Centralized management simplifies deployment, updates, and monitoring, streamlining system administration \cite{wang2023shoggoth}. Consequently, cloud processing allows edge and terminal devices to focus on data acquisition and preprocessing, while the cloud handles model training and complex inference, optimizing resource utilization across the system \cite{lu2024samedge}.
\subsection{Training Strategies}\label{sec4.4}
In this section, we review training strategies for cloud-centric video analytics, including supervised, unsupervised, and transfer learning.
\paragraph{Supervised Learning.}\label{sec4.4.1}
Deep learning models have been widely adopted in cloud-centric video analytics, demonstrating outstanding performance in pattern and behavior recognition, with capabilities in accurate anomaly detection and behavior analysis \cite{xu2024unleashing}. Cloud infrastructure enables training of complex CNNs that learn spatio-temporal representations, which can be deployed across CETC systems for tasks ranging from ego-vehicle action recognition to unintentional action identification \cite{noguchi2023egovehicle}. Additionally, model performance relies on representation diversity and feature transferability, with techniques like variance-covariance regularization (VCReg) promoting high-variance, low-covariance representations across diverse tasks \cite{zhu2024variancecovariance}. Moreover, in open-set video anomaly detection, where models need to generalize to unseen anomaly types, the choice of training data becomes crucial for ensuring robust performance, while the cost of labeling diverse anomaly types can be substantial \cite{zhu2022open}.

\paragraph{Unsupervised Learning.}\label{sec4.4.2}
Novel pattern discovery and anomaly identification emerge through unlabeled video data analysis, offering particular value when labeled data proves scarce or costly \cite{liu2024generalized}. For example, deep probabilistic models and clustering techniques like GMM-DAE \cite{ouyang2021video}, enable outlier detection in latent feature spaces. Recently approaches integrate evidential deep learning with normalizing flows \cite{zhu2022open} to address open-set video anomaly detection, while physical law discovery from distorted video leverages autoencoders and symbolic regression \cite{udrescu2021symbolic}. DyAnNet \cite{thakare2023dyannet} uses unsupervised clustering and scene dynamicity to generate preliminary anomaly scores, subsequently refined by a cross-branch feed-forward network.  Similarly, structural feature-autoencoders \cite{meissen2023unsupervised} address imperfect reconstruction in medical images by transforming input images into a feature space for anomaly detection along different discriminative feature maps.
\paragraph{Transfer Learning.}\label{sec4.4.3}
Pre-trained model adaptation leverages edge-cloud collaborative architectures to enhance performance and efficiency in video analytics tasks, utilizing distributed computational resources and optimized model deployment strategies \cite{xu2024unleashing}. The success of transfer learning in video analytics is strongly influenced by the alignment between the source and target domains. For example, Video-LaVIT \cite{jin2024videolavit} leverages unified visual-motional tokenization, and has shown promising results in transferring knowledge across diverse video-language tasks. Transferring knowledge from action recognition in generic videos to recognizing surgical procedures in medical videos could be effective due to shared features like human motion, while vastly different source and target tasks may hinder performance \cite{caccia2023computeoptimal}. In addition, the size and quality of target dataset also influence fine-tuning outcomes, while techniques like co-finetuning enhance rare class performance \cite{arnab2022transfer}. Similarly, variance-covariance regularization (VCReg) \cite{zhu2024variancecovariance} promotes feature diversity, complemented by structured pruning approaches that optimize computational efficiency.

\subsection{Inference Optimization}\label{sec4.5}
Emerging cloud-based inference optimization techniques address the challenges of large-scale video analytics through parallel processing, distributed computation, and efficient model serving strategies.
\paragraph{Model Parallelism.}\label{sec4.5.1}
Distributed computation across multiple cloud instances enables efficient execution of large-scale video analytics models through parallel processing, particularly benefiting complex architectures that exceed single-machine capabilities~\cite{tian2024dynamic}. It benefits complex video models that are too large for single machines or require excessive inference times. For example, massively parallel video networks (MPVN) \cite{carreira2018massively} use operation pipelining and multi-rate clocks for parallel computation across multiple GPUs, minimizing latency and maximizing throughput for tasks like action recognition. Similarly, CROSSBOW \cite{koliousis2019crossbow} employs synchronous model averaging (SMA) to improve training time by allowing replicas to explore the solution space independently while maintaining consistency, and systems like iGniter \cite{xu2023igniter} address GPU sharing challenges through proactive, interference-aware resource provisioning for predictable DNN inference and cost-efficient allocation.

\paragraph{Batch Processing.}\label{sec4.5.2}
Concurrent processing of multiple video streams enables significant efficiency gains through cloud-based computational resources, optimizing both processing time and cost-effectiveness compared to sequential approaches. For example, platforms like Llama~\cite{romero2021llama} leverage heterogeneous serverless frameworks for dynamic resource allocation, while systems such as MPVN \cite{carreira2018massively} integrate model parallelism techniques (e.g., VPaaS \cite{zhang2021serverless}) to maximize throughput and minimize latency. However, effective batch processing requires careful resource allocation and scheduling to avoid bottlenecks, which analyze aggregating functions in serverless clouds for video streaming \cite{wu2020descriptive}.

\paragraph{Model Serving.}\label{sec4.5.3}

Deploying video analytics models in the cloud requires accessible APIs for applications and users. Cloud platforms provide robust infrastructure for efficient model management, scaling, and monitoring, ensuring high availability through dynamic resource allocation. Tools like TensorFlow Serving and AWS SageMaker enable seamless deployment and updates of models, facilitating reliable service delivery \cite{vo2022edge}. For example, Hyper~\cite{buniatyan2019hyper} offers a hybrid distributed cloud solution for large-scale deep learning tasks, encompassing both model deployment and inference. Additionally, model quantization like HQ \cite{liu2023hyperspherical} can reduce model size and improve accuracy, leading to more efficient model serving.  In addition, CROSSBOW \cite{koliousis2019crossbow}, a system for training deep learning models on multi-GPU servers, has the potential to be extended for optimized model serving in similar cloud environments.
\section{Hybrid Video Analytics}\label{sec5}
\begin{figure}[t]
    \centering
    \includegraphics[width=0.48\textwidth]{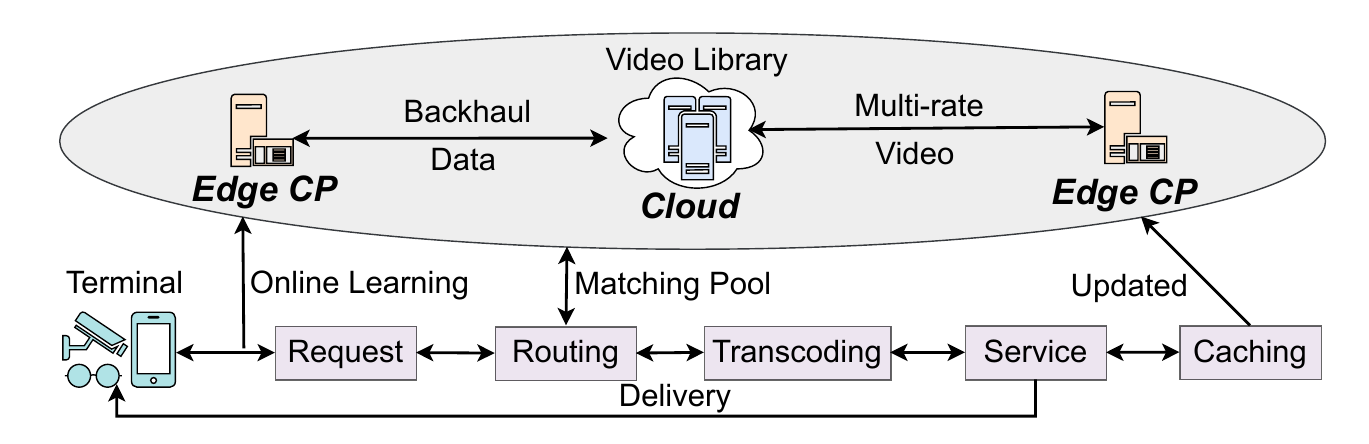}
    \vspace{-20px}
    \caption{Example of CETC video analytics workflow.}
    \label{fig:workflow}
    \vspace{-12pt}
\end{figure}
\subsection{Adaptive Task Offloading}\label{sec5.1}
Dynamic workload distribution across CETC video analytics systems responds intelligently to network bandwidth fluctuations, resource availability, and application demands \cite{tian2024dynamic}. Under constrained network conditions, latency-sensitive operations migrate to edge nodes to circumvent transmission bottlenecks \cite{silva2021energyaware}, whereas cloud resources activate when edge capacity thresholds are exceeded. The Shoggoth architecture \cite{wang2023shoggoth} dynamically adjusts task allocation between edge and cloud based on these factors, optimizing inference performance. Contemporary research diverges into multiple optimization paradigms: token bucket regulated task scheduling in ECIVA \cite{ji2024task} contrasts with hierarchical approaches like dual-timescale service caching \cite{li2024twotimescale} and meta-reinforcement learning architectures, DMQTO \cite{sharma2024deep}, leverage online and reinforcement learning to optimize offloading decisions. Building upon these foundations, task dependency modeling in human digital twin systems, ECAO \cite{zhang2024energy}, introduces context-aware orchestration, while transcoding-aware caching strategies \cite{xiao2024transcodingenabled} optimize content delivery through adaptive bitrate adjustments, as illustrated in Fig.~\ref{fig:workflow}.

\subsection{Resource-aware Scheduling}\label{sec5.2}
Resource-aware scheduling intelligently distributes video processing tasks across the heterogeneous cloud-edge-terminal infrastructure, considering diverse capabilities and constraints to optimize system performance \cite{zhang2024hybrid}.  Specifically, these approaches aim to minimize latency and energy consumption while maximizing resource utilization and meeting quality-of-service requirements. For example, computationally intensive tasks can be offloaded to the resource-rich cloud, while latency-sensitive tasks can be executed on edge devices \cite{ji2024task}.  Several key works address this challenge. \cite{li2024twotimescale} propose a two-timescale framework for joint service caching and resource allocation, using Lyapunov-based optimization to maximize long-term network performance under energy constraints.  In a different vein,~\cite{lu2024a2cdrl} introduce A2C-DRL, a dynamic scheduling technique leveraging Advantage Actor-Critic and deep reinforcement learning for efficient resource utilization and load balancing in stochastic edge-cloud environments. Additionally, EdgeVision \cite{gao2024edgevision} employs multiagent reinforcement learning for resource-aware scheduling, enabling autonomous edge nodes to learn optimal policies for workload distribution and request dispatching. Similarly,~\cite{zhang2024hybrid} combines deep reinforcement learning and multi-objective optimization, considering task completion time, cost, energy consumption, and system reliability for balanced performance.
\section{Challenges and Future Directions}\label{sec6}
Despite significant progress in video analytics, challenges persist in CETC systems, and we identify and discuss key issues and future directions in this section.
\paragraph{Platform Integration.}\label{sec6.1}
Heterogeneous hardware and diverse software environments pose significant challenges to seamless interoperability in CETC video analytics systems.  Specifically, achieving efficient data exchange, consistent model deployment, and coordinated task execution across these disparate platforms introduces complexities. To address these challenges, standardized interfaces and APIs can abstract hardware and software differences, enabling portable and efficient code development.  In addition, containerization and virtualization technologies simplify deployment and management across diverse platforms, while optimized communication protocols and data serialization formats minimize overhead and maximize performance.
\paragraph{System Scalability.}\label{sec6.2}
Edge video analytics systems face significant scalability challenges due to the growing number of edge devices and video streams, where distributed collaboration complexities and device heterogeneity introduce communication bottlenecks and inefficient resource allocation \cite{vo2022edge}. To address these challenges, decentralized management strategies via MARL enable autonomous resource allocation and task offloading \cite{gao2024edgevision}, while efficient communication protocols minimize overhead through hierarchical or peer-to-peer sharing \cite{ji2024task}. Furthermore, optimizing data processing pipelines via adaptive online learning \cite{wang2023shoggoth,lu2023turbo}, task-specific discrimination, and incorporating active inference mechanisms \cite{sedlak2024equilibrium} further enhance system scalability.
\paragraph{Data Protection.}\label{sec6.3}
Privacy and security considerations in distributed cloud-edge-terminal environments demand robust protection mechanisms for sensitive video data. To address this, federated learning emerges as a key solution, enabling collaborative model training without raw data sharing through frameworks like Fed-FSNet~\cite{guo2023fedfsnet}. For better data protection, HierSFL~\cite{quan2023hiersfl} incorporates local differential privacy with noise-added updates. Device heterogeneity poses significant challenges to efficiency and fairness in these systems. To address these issues, fuzzy networks facilitate synthetic i.i.d. data generation to mitigate distribution biases. Additionally, complementary approaches such as secure multi-party computation, homomorphic encryption, and differential privacy provide additional safeguards despite computational overhead~\cite{guo2023fedfsnet}.

\paragraph{System Resilience.}\label{sec6.4}
Video analytics in CETC systems require robust fault tolerance due to potential disruptions like network failures and device outages. For example, load rebalancing within a device cluster can restore individual edge device performance after network failures \cite{sedlak2024equilibrium}. The dynamic nature of these environments necessitates sophisticated adaptation mechanisms, particularly in task offloading frameworks that leverage enhanced deep Q-networks (EDQN) \cite{ji2024task} to optimize resource allocation during system failures. In addtion, resource-aware scheduling \cite{gao2024edgevision} further bolsters resilience through dynamic resource allocation based on real-time conditions. Redundancy in hardware and software, including backup servers and data replication, and blockchain technology for secure index authentication \cite{nikouei2018realtime} also enhances resilience. Additionally, graceful degradation strategies, which prioritize essential tasks and reduce the quality of non-critical processes during disruptions, ensure a baseline service level.
\paragraph{Explainable and Efficient Systems.}\label{sec6.5}
Explainable video analytics models enhance trust and transparency by providing insights into their decision-making processes, while efficient system design and specialized hardware accelerate edge processing through techniques like heterogeneous computing-based video segmentation~\cite{lin2022efficient} and turbo opportunistic enhancement~\cite{lu2023turbo} for latency-preserving accuracy improvements. Emerging large language model-multi-agent (LLM-MA) systems~\cite{ning2023videobench} advance video understanding and content generation~\cite{zhou2024survey} by combining multiple modalities~\cite{lu2024samedge}, though platform integration and scalability challenges~\cite{romero2021llama} must be addressed to realize the full potential of efficient and explainable LLM-MA systems in collaborative environments~\cite{bhardwaj2022ekya}.

\paragraph{Advanced Video Analytics.}\label{sec6.6}
Recent advances in video analytics have been revolutionized by LLMs and multimodal integration approaches. Video-LLMs and Video-LLaVA demonstrate sophisticated video understanding capabilities, while specialized benchmarks (e.g., Video-Bench and VLM-Eval) enable systematic evaluation~\cite{ning2023videobench}. Resource-efficient solutions leverage image LLM priors, and advanced architectures such as Video-LaVIT~\cite{jin2024videolavit} employ decoupled visual-motional tokenization for unified video-language pre-training. Integrating video data with multimodal integration of audio and text and enhanced time series analysis capabilities, offer opportunities for comprehensive video understanding and analysis~\cite{zhou2024survey}.
\vspace{-6pt}
\section{Conclusion}\label{sec7}
In this survey, we present an overview of video analytics in CETC systems, covering architectural paradigms, key technologies, and optimization strategies. Moreover, we highlight the benefits and challenges of edge-centric and cloud-centric approaches, emphasizing the importance of hybrid methods. Despite significant progress, challenges in platform integration, system scalability, data protection, and system resilience still require further research and innovative solutions.
\bibliographystyle{named}
\begin{spacing}{0.82} 
\bibliography{IJCAI2025.bib}

\begin{thebibliography}{}

\bibitem[\protect\citeauthoryear{Arnab \bgroup \em et al.\egroup }{2022}]{arnab2022transfer}
Anurag Arnab, Xuehan Xiong, Alexey Gritsenko, Rob Romijnders, Josip Djolonga, Mostafa Dehghani, Chen Sun, Mario Lu{\v c}i{\'c}, and Cordelia Schmid.
\newblock Beyond transfer learning: Co-finetuning for action localisation, 2022.

\bibitem[\protect\citeauthoryear{Bayhan \bgroup \em et al.\egroup }{2021}]{bayhan2021edgedash}
Suzan Bayhan, Setareh Maghsudi, and Anatolij Zubow.
\newblock {EdgeDASH}: Exploiting network-assisted adaptive video streaming for edge caching.
\newblock {\em IEEE TNSM}, 18(2):1732--1745, 2021.

\bibitem[\protect\citeauthoryear{Bhardwaj \bgroup \em et al.\egroup }{2022}]{bhardwaj2022ekya}
Romil Bhardwaj, Zhengxu Xia, Ganesh Ananthanarayanan, Junchen Jiang, Yuanchao Shu, Nikolaos Karianakis, Kevin Hsieh, Paramvir Bahl, and Ion Stoica.
\newblock Ekya: Continuous learning of video analytics models on edge compute servers.
\newblock In {\em NSDI}, 2022.

\bibitem[\protect\citeauthoryear{Buniatyan}{2019}]{buniatyan2019hyper}
Davit Buniatyan.
\newblock Hyper: Distributed cloud processing for large-scale deep learning tasks.
\newblock In {\em CSIT}, pages 27--32, 2019.

\bibitem[\protect\citeauthoryear{Caccia \bgroup \em et al.\egroup }{2023}]{caccia2023computeoptimal}
Massimo Caccia, Alexandre Galashov, Arthur Douillard, Amal {Rannen-Triki}, Dushyant Rao, Michela Paganini, Laurent Charlin, Marc'Aurelio Ranzato, and Razvan Pascanu.
\newblock Towards compute-optimal transfer learning.
\newblock {\em arXiv:2304.13164}, 2023.

\bibitem[\protect\citeauthoryear{Carreira \bgroup \em et al.\egroup }{2018}]{carreira2018massively}
Jo{\~a}o Carreira, Viorica P{\u a}tr{\u a}ucean, Laurent Mazare, Andrew Zisserman, and Simon Osindero.
\newblock Massively parallel video networks.
\newblock In {\em ECCV}, 2018.

\bibitem[\protect\citeauthoryear{Fan \bgroup \em et al.\egroup }{2018}]{fan2018endtoend}
Lijie Fan, Wenbing Huang, Chuang Gan, Stefano Ermon, Boqing Gong, and Junzhou Huang.
\newblock End-to-end learning of motion representation for video understanding.
\newblock In {\em CVPR}, pages 6016--6025, 2018.

\bibitem[\protect\citeauthoryear{Gao \bgroup \em et al.\egroup }{2024}]{gao2024edgevision}
Guanyu Gao, Yuqi Dong, Ran Wang, and Xin Zhou.
\newblock {EdgeVision}: Towards collaborative video analytics on distributed edges for performance maximization.
\newblock {\em IEEE TMM}, 26:9083--9094, 2024.

\bibitem[\protect\citeauthoryear{Gu \bgroup \em et al.\egroup }{2024}]{gu2024aienhanced}
Huixian Gu, Liqiang Zhao, Zhu Han, Gan Zheng, and Shenghui Song.
\newblock {AI}-enhanced cloud-edge-terminal collaborative network: Survey, applications, and future directions.
\newblock {\em IEEE COMST}, 26(2):1322--1385, 2024.

\bibitem[\protect\citeauthoryear{Guerra and Drummond}{2021}]{guerra2021automatic}
Luis Guerra and Tom Drummond.
\newblock Automatic pruning for quantized neural networks.
\newblock In {\em DICTA}, pages 01--08, 2021.

\bibitem[\protect\citeauthoryear{Guo \bgroup \em et al.\egroup }{2023}]{guo2023fedfsnet}
Jingcai Guo, Song Guo, Jie Zhang, and Ziming Liu.
\newblock {Fed-FSNet}: Mitigating non-i.i.d. federated learning via fuzzy synthesizing network, 2023.

\bibitem[\protect\citeauthoryear{Hua \bgroup \em et al.\egroup }{2024}]{hua2024energyefficient}
Wei Hua, Peng Liu, and Linyu Huang.
\newblock Energy-efficient resource allocation for heterogeneous edge--cloud computing.
\newblock {\em IEEE IoTJ}, 11(2), 2024.

\bibitem[\protect\citeauthoryear{Ji \bgroup \em et al.\egroup }{2024}]{ji2024task}
Xiaofeng Ji, Faming Gong, Nuanlai Wang, Chengze Du, and Xiangbing Yuan.
\newblock Task offloading with enhanced deep q-networks for efficient industrial intelligent video analysis in edge--cloud collaboration.
\newblock {\em Adv. Eng. Inform.}, 62:102599, 2024.

\bibitem[\protect\citeauthoryear{Jin \bgroup \em et al.\egroup }{2024}]{jin2024videolavit}
Yang Jin, Zhicheng Sun, Kun Xu, Kun Xu, Liwei Chen, Hao Jiang, Quzhe Huang, Chengru Song, Yuliang Liu, Di~Zhang, Yang Song, Kun Gai, and Yadong Mu.
\newblock Video-lavit: Unified video-language pre-training with decoupled visual-motional tokenization, 2024.

\bibitem[\protect\citeauthoryear{Joshi \bgroup \em et al.\egroup }{2024}]{joshi2024integration}
Amrita Joshi, Saurabh Agarwal, Debi~Prasanna Kanungo, and Rajib~Kumar Panigrahi.
\newblock Integration of edge--ai into iot--cloud architecture for landslide monitoring and prediction.
\newblock {\em IEEE TII}, 20(3):4246--4258, 2024.

\bibitem[\protect\citeauthoryear{Koliousis \bgroup \em et al.\egroup }{2019}]{koliousis2019crossbow}
Alexandros Koliousis, Pijika Watcharapichat, Matthias Weidlich, Luo Mai, Paolo Costa, and Peter Pietzuch.
\newblock {CROSSBOW}: Scaling deep learning with small batch sizes on multi-gpu servers.
\newblock {\em PVLDB}, 12(11):1399--1412, 2019.

\bibitem[\protect\citeauthoryear{Kuswiradyo \bgroup \em et al.\egroup }{2024}]{kuswiradyo2024optimizing}
Primatar Kuswiradyo, Binayak Kar, and Shan-Hsiang Shen.
\newblock Optimizing the energy consumption in three-tier cloud--edge--fog federated systems with omnidirectional offloading.
\newblock {\em CN}, 250:110578, 2024.

\bibitem[\protect\citeauthoryear{Li \bgroup \em et al.\egroup }{2024}]{li2024twotimescale}
Yafei Li, Huiqiang Wang, Jiayu Sun, Hongwu Lv, Wenqi Zheng, and Guangsheng Feng.
\newblock Two-timescale joint service caching and resource allocation for task offloading with edge--cloud cooperation.
\newblock {\em CN}, 254:110771, 2024.

\bibitem[\protect\citeauthoryear{Lin \bgroup \em et al.\egroup }{2022}]{lin2022efficient}
Jamie~Menjay Lin, Siargey Pisarchyk, Juhyun Lee, David Tian, Tingbo Hou, Karthik Raveendran, Raman Sarokin, George Sung, Trent Tolley, and Matthias Grundmann.
\newblock Efficient heterogeneous video segmentation at the edge.
\newblock {\em arXiv:2208.11666}, 2022.

\bibitem[\protect\citeauthoryear{Liu \bgroup \em et al.\egroup }{2023a}]{liu2023hyperspherical}
Dan Liu, Xi~Chen, Chen Ma, and Xue Liu.
\newblock Hyperspherical quantization: Toward smaller and more accurate models.
\newblock In {\em WACV}, pages 5251--5261, 2023.

\bibitem[\protect\citeauthoryear{Liu \bgroup \em et al.\egroup }{2023b}]{liu2023singlepath}
Jing Liu, Bohan Zhuang, Peng Chen, Chunhua Shen, Jianfei Cai, and Mingkui Tan.
\newblock Single-path bit sharing for automatic loss-aware model compression.
\newblock {\em IEEE TPAMI}, 45(10):12459--12473, 2023.

\bibitem[\protect\citeauthoryear{Liu \bgroup \em et al.\egroup }{2024}]{liu2024generalized}
Yang Liu, Dingkang Yang, Yan Wang, Jing Liu, Jun Liu, Azzedine Boukerche, Peng Sun, and Liang Song.
\newblock Generalized video anomaly event detection: Systematic taxonomy and comparison of deep models.
\newblock {\em ACM CSUR}, 2024.

\bibitem[\protect\citeauthoryear{Lu \bgroup \em et al.\egroup }{2023a}]{lu2023turbo}
Yan Lu, Shiqi Jiang, Ting Cao, and Yuanchao Shu.
\newblock Turbo: Opportunistic enhancement for edge video analytics.
\newblock In {\em SenSys}, pages 263--276, 2023.

\bibitem[\protect\citeauthoryear{Lu \bgroup \em et al.\egroup }{2023b}]{lu2023transflow}
Yawen Lu, Qifan Wang, Siqi Ma, Tong Geng, Yingjie~Victor Chen, Huaijin Chen, and Dongfang Liu.
\newblock {TransFlow}: Transformer as flow learner.
\newblock In {\em CVPR}, pages 18063--18073, 2023.

\bibitem[\protect\citeauthoryear{Lu \bgroup \em et al.\egroup }{2024a}]{lu2024a2cdrl}
Jialin Lu, Jing Yang, Shaobo Li, Yijun Li, Wu~Jiang, Jiangtian Dai, and Jianjun Hu.
\newblock A2c-drl: Dynamic scheduling for stochastic edge--cloud environments using a2c and deep reinforcement learning.
\newblock {\em IEEE IoTJ}, 11(9):16915--16927, 2024.

\bibitem[\protect\citeauthoryear{Lu \bgroup \em et al.\egroup }{2024b}]{lu2024samedge}
Rui Lu, Siping Shi, Yanting Liu, and Dan Wang.
\newblock Samedge: An edge-cloud video analytics architecture for the segment anything model, 2024.

\bibitem[\protect\citeauthoryear{Meissen \bgroup \em et al.\egroup }{2023}]{meissen2023unsupervised}
Felix Meissen, Johannes Paetzold, Georgios Kaissis, and Daniel Rueckert.
\newblock Unsupervised anomaly localization with~structural feature-autoencoders.
\newblock In {\em Brainlesion}, pages 14--24, 2023.

\bibitem[\protect\citeauthoryear{Neff \bgroup \em et al.\egroup }{2020}]{neff2020revamp2t}
Christopher Neff, Mat{\'i}as Mendieta, Shrey Mohan, Mohammadreza Baharani, Samuel Rogers, and Hamed Tabkhi.
\newblock Revamp2t: Real-time edge video analytics for multicamera privacy-aware pedestrian tracking.
\newblock {\em IEEE IoTJ}, 7(4):2591--2602, 2020.

\bibitem[\protect\citeauthoryear{Nikouei \bgroup \em et al.\egroup }{2018}]{nikouei2018realtime}
Seyed~Yahya Nikouei, Ronghua Xu, Deeraj Nagothu, Yu~Chen, Alexander Aved, and Erik Blasch.
\newblock Real-time index authentication for event-oriented surveillance video query using blockchain.
\newblock In {\em ISC2}, 2018.

\bibitem[\protect\citeauthoryear{Ning \bgroup \em et al.\egroup }{2023}]{ning2023videobench}
Munan Ning, Bin Zhu, Yujia Xie, Bin Lin, Jiaxi Cui, Lu~Yuan, Dongdong Chen, and Li~Yuan.
\newblock Video-bench: A comprehensive benchmark and toolkit for evaluating video-based large language models, 2023.

\bibitem[\protect\citeauthoryear{Noguchi and Tanizawa}{2023}]{noguchi2023egovehicle}
Chihiro Noguchi and Toshihiro Tanizawa.
\newblock Ego-vehicle action recognition based on semi-supervised contrastive learning.
\newblock In {\em WACV}, pages 5977--5987, 2023.

\bibitem[\protect\citeauthoryear{Ollivier \bgroup \em et al.\egroup }{2023}]{ollivier2023sustainable}
S{\'e}bastien Ollivier, Sheng Li, Yue Tang, Stephen Cahoon, Ryan Caginalp, Chayanika Chaudhuri, Peipei Zhou, Xulong Tang, Jingtong Hu, and Alex~K. Jones.
\newblock Sustainable ai processing at the edge.
\newblock {\em IEEE Micro}, 43(1):19--28, 2023.

\bibitem[\protect\citeauthoryear{Ouyang and Sanchez}{2021}]{ouyang2021video}
Yuqi Ouyang and Victor Sanchez.
\newblock Video anomaly detection by estimating likelihood of representations.
\newblock In {\em ICPR}, 2021.

\bibitem[\protect\citeauthoryear{Quan \bgroup \em et al.\egroup }{2023}]{quan2023hiersfl}
Minh~K. Quan, Dinh~C. Nguyen, Van-Dinh Nguyen, Mayuri Wijayasundara, Sujeeva Setunge, and Pubudu~N. Pathirana.
\newblock {HierSFL}: Local differential privacy-aided split federated learning in mobile edge computing.
\newblock In {\em VCC}, pages 103--108, 2023.

\bibitem[\protect\citeauthoryear{Rajagopal and Bouganis}{2021}]{rajagopal2021perf4sight}
Aditya Rajagopal and Christos-Savvas Bouganis.
\newblock Perf4sight: A toolflow to model cnn training performance on edge gpus.
\newblock In {\em ICCVW}, pages 963--971, 2021.

\bibitem[\protect\citeauthoryear{Romero \bgroup \em et al.\egroup }{2021}]{romero2021llama}
Francisco Romero, Mark Zhao, Neeraja~J. Yadwadkar, and Christos Kozyrakis.
\newblock Llama: A heterogeneous \& serverless framework for auto-tuning video analytics pipelines.
\newblock In {\em SoCC}, pages 1--17, 2021.

\bibitem[\protect\citeauthoryear{Sedlak \bgroup \em et al.\egroup }{2024}]{sedlak2024equilibrium}
Boris Sedlak, Victor~Casamayor Pujol, Praveen~Kumar Donta, and Schahram Dustdar.
\newblock Equilibrium in the computing continuum through active inference.
\newblock {\em FGCS}, 160:92--108, 2024.

\bibitem[\protect\citeauthoryear{Sharma \bgroup \em et al.\egroup }{2024}]{sharma2024deep}
Nelson Sharma, Aswini Ghosh, Rajiv Misra, and Sajal~K. Das.
\newblock Deep meta q-learning based multi-task offloading in edge-cloud systems.
\newblock {\em IEEE TMC}, 23(4):2583--2598, 2024.

\bibitem[\protect\citeauthoryear{Silva \bgroup \em et al.\egroup }{2021}]{silva2021energyaware}
Joaquim Silva, Eduardo R.~B. Marques, Lu{\'i}s~M.B. Lopes, and Fernando Silva.
\newblock Energy-aware adaptive offloading of soft real-time jobs in mobile edge clouds.
\newblock {\em J. Cloud Comput.}, 10(1):38, 2021.

\bibitem[\protect\citeauthoryear{Stephenson \bgroup \em et al.\egroup }{2019}]{stephenson2019degrafflow}
Felix Stephenson, Toby~P. Breckon, and Ioannis Katramados.
\newblock Degraf-flow: Extending degraf features for accurate and efficient sparse-to-dense optical flow estimation.
\newblock In {\em ICIP}, pages 1277--1281, 2019.

\bibitem[\protect\citeauthoryear{Thakare \bgroup \em et al.\egroup }{2023}]{thakare2023dyannet}
Kamalakar~Vijay Thakare, Yash Raghuwanshi, Debi~Prosad Dogra, Heeseung Choi, and Ig-Jae Kim.
\newblock Dyannet: A scene dynamicity guided self-trained video anomaly detection network.
\newblock In {\em WACV}, pages 5530--5539, 2023.

\bibitem[\protect\citeauthoryear{Tian \bgroup \em et al.\egroup }{2024}]{tian2024dynamic}
Xianzhong Tian, Huixiao Meng, Yifan Shen, Junxian Zhang, Yuzhe Chen, and Yanjun Li.
\newblock Dynamic microservice deployment and offloading for things--edge--cloud computing.
\newblock {\em IEEE IoTJ}, 11(11):1--11, 2024.

\bibitem[\protect\citeauthoryear{Tran \bgroup \em et al.\egroup }{2024}]{tran2024deeplearning}
Phu~N. Tran, Sattvic Ray, Linnea Lemma, Yunrui Li, Reef Sweeney, Aparna Baskaran, Zvonimir Dogic, Pengyu Hong, and Michael~F. Hagan.
\newblock Deep-learning optical flow outperforms piv in obtaining velocity fields from active nematics.
\newblock {\em arXiv:2404.15497}, 2024.

\bibitem[\protect\citeauthoryear{Udrescu and Tegmark}{2021}]{udrescu2021symbolic}
Silviu-Marian Udrescu and Max Tegmark.
\newblock Symbolic pregression: Discovering physical laws from distorted video.
\newblock {\em Phys. Rev. E}, 103(4):043307, 2021.

\bibitem[\protect\citeauthoryear{Vo \bgroup \em et al.\egroup }{2022}]{vo2022edge}
Thong Vo, Pranjal Dave, Gaurav Bajpai, and Rasha Kashef.
\newblock Edge, fog, and cloud computing : An overview on challenges and applications, 2022.

\bibitem[\protect\citeauthoryear{Wang \bgroup \em et al.\egroup }{2023}]{wang2023shoggoth}
Liang Wang, Kai Lu, Nan Zhang, Xiaoyang Qu, Jianzong Wang, Jiguang Wan, Guokuan Li, and Jing Xiao.
\newblock Shoggoth: Towards efficient edge-cloud collaborative real-time video inference via adaptive online learning.
\newblock In {\em DAC}, pages 1--6, 2023.

\bibitem[\protect\citeauthoryear{Wu \bgroup \em et al.\egroup }{2020}]{wu2020descriptive}
Shangrui Wu, Chavit Denninnart, Xiangbo Li, Yang Wang, and Mohsen~Amini Salehi.
\newblock Descriptive and predictive analysis of aggregating functions in serverless clouds: The case of video streaming.
\newblock In {\em HPCC}, 2020.

\bibitem[\protect\citeauthoryear{Xiao \bgroup \em et al.\egroup }{2024}]{xiao2024transcodingenabled}
Han Xiao, Yirong Zhuang, Changqiao Xu, Wendong Wang, Hongke Zhang, Renjie Ding, Tengfei Cao, Lujie Zhong, and Gabriel-Miro Muntean.
\newblock Transcoding-enabled cloud--edge--terminal collaborative video caching in heterogeneous iot networks: An online learning approach with time-varying information.
\newblock {\em IEEE IoTJ}, 11(1):296--310, 2024.

\bibitem[\protect\citeauthoryear{Xu \bgroup \em et al.\egroup }{2019}]{xu2019robust}
Zhe Xu, Biao Min, and Ray~C.C. Cheung.
\newblock A robust background initialization algorithm with superpixel motion detection.
\newblock {\em Signal Process.: Image Commun.}, 71:1--12, 2019.

\bibitem[\protect\citeauthoryear{Xu \bgroup \em et al.\egroup }{2023}]{xu2023igniter}
Fei Xu, Jianian Xu, Jiabin Chen, Li~Chen, Ruitao Shang, Zhi Zhou, and Fangming Liu.
\newblock igniter: Interference-aware gpu resource provisioning for predictable dnn inference in the cloud.
\newblock {\em IEEE TPDS}, 34(3):812--827, 2023.

\bibitem[\protect\citeauthoryear{Xu \bgroup \em et al.\egroup }{2024}]{xu2024unleashing}
Minrui Xu, Hongyang Du, Dusit Niyato, Jiawen Kang, Zehui Xiong, Shiwen Mao, Zhu Han, Abbas Jamalipour, Dong~In Kim, Xuemin Shen, Victor C.~M. Leung, and H.~Vincent Poor.
\newblock Unleashing the power of edge-cloud generative ai in mobile networks: A survey of aigc services.
\newblock {\em IEEE COMST}, 26(2):1127--1170, 2024.

\bibitem[\protect\citeauthoryear{Xue \bgroup \em et al.\egroup }{2021}]{xue2021denoisingbased}
Zhipeng Xue, Xiaojun Yuan, and Yang Yang.
\newblock Denoising-based turbo message passing for compressed video background subtraction.
\newblock {\em IEEE TIP}, 30:2682--2696, 2021.

\bibitem[\protect\citeauthoryear{Zhang \bgroup \em et al.\egroup }{2021}]{zhang2021serverless}
Huaizheng Zhang, Meng Shen, Yizheng Huang, Yonggang Wen, Yong Luo, Guanyu Gao, and Kyle Guan.
\newblock A serverless cloud-fog platform for dnn-based video analytics with incremental learning.
\newblock {\em arXiv:2102.03012}, 2021.

\bibitem[\protect\citeauthoryear{Zhang \bgroup \em et al.\egroup }{2024a}]{zhang2024hybrid}
Jiangjiang Zhang, Zhenhu Ning, Muhammad Waqas, and Sheng Chen.
\newblock Hybrid edge-cloud collaborator resource scheduling approach based on deep reinforcement learning and multiobjective optimization.
\newblock {\em IEEE TC}, 73(1):192--205, 2024.

\bibitem[\protect\citeauthoryear{Zhang \bgroup \em et al.\egroup }{2024b}]{zhang2024energy}
Qiang Zhang, Yuye Yang, Changyan Yi, Samuel~D. Okegbile, and Jun Cai.
\newblock Energy- and cost-aware offloading of dependent tasks with edge-cloud collaboration for human digital twin.
\newblock {\em IEEE IoTJ}, 11(17):29116--29131, 2024.

\bibitem[\protect\citeauthoryear{Zhou \bgroup \em et al.\egroup }{2024}]{zhou2024survey}
Pengyuan Zhou, Lin Wang, Zhi Liu, Yanbin Hao, Pan Hui, Sasu Tarkoma, and Jussi Kangasharju.
\newblock A survey on generative ai and llm for video generation, understanding, and streaming, 2024.

\bibitem[\protect\citeauthoryear{Zhu \bgroup \em et al.\egroup }{2022}]{zhu2022open}
Yuansheng Zhu, Wentao Bao, and Qi~Yu.
\newblock Towards open set video anomaly detection.
\newblock In {\em ECCV}, pages 395--412, 2022.

\bibitem[\protect\citeauthoryear{Zhu \bgroup \em et al.\egroup }{2024a}]{zhu2024variancecovariance}
Jiachen Zhu, Katrina Evtimova, Yubei Chen, Ravid {Shwartz-Ziv}, and Yann LeCun.
\newblock Variance-covariance regularization improves representation learning.
\newblock {\em arXiv:2306.13292}, 2024.

\bibitem[\protect\citeauthoryear{Zhu \bgroup \em et al.\egroup }{2024b}]{zhu2024edge}
Jing Zhu, Chuanjiang Hu, Edris Khezri, and Mohd Mustafa~Mohd Ghazali.
\newblock Edge intelligence-assisted animation design with large models: A survey.
\newblock {\em J. Cloud Comput.}, 13(1):48, 2024.

\end{thebibliography}
\end{spacing}
\end{document}